\documentclass[11pt]{article}

\usepackage{graphicx}

\begin{document}

\pagestyle{empty}

\renewcommand{\thefootnote}{\fnsymbol{footnote}}


\vspace{8mm}

\begin{center}
{\Large\bf
ILC Extraction Line Simulations with TDR Parameters\footnote
{Work supported by the US Department of Energy contract
DE--AC02--76SF00515.
}}

\vspace{1cm}

\large{
E.~Mar{\'\i}n and Y.~Nosochkov\\
SLAC National Accelerator Laboratory, Menlo Park, CA 94025
}

\end{center}

\vfill

\begin{center}
{\large\bf
Abstract
}
\end{center}

\begin{quote}
\large{

The goal of this study is to evaluate the impact of the latest ILC beam
parameters at the Interaction Point (IP), as specified in the 2013 ILC
Technical Design Report (TDR), on beam losses in the extraction
line. The previous beam loss evaluation was based on the
parameters specified in the 2007 ILC Reference Design Report
(RDR). The results of this study are compared to the results obtained
in the past
for the ``nominal'' and the ``low power'' (low-P) parameter options of
the RDR. The initial disrupted beam distribution at IP was generated
using Guinea-Pig code, and the beam losses were obtained in tracking
simulations using DIMAD. The study is performed for 500 GeV
center-of-mass beam energy and the extraction line optics
corresponding to the latest final focus optics with $L^*=4.5$ m,
with and without detector solenoid.
}
\end{quote}

\vfill

\begin{center}
\large{
\itshape{
Talk presented at the International Workshop on Future Linear Colliders
(LCWS13)\\
Tokyo, Japan, 11-15 November 2013
}}
\end{center}

\newpage

\pagenumbering{arabic}
\pagestyle{plain}

\vspace{-0mm}
\section{Introduction}

The ILC extraction line \cite{exline} is designed for 14 mrad
horizontal crossing angle at the Interaction Point (IP).
The extraction optics provides large beam acceptance
in order to minimize beam losses caused by long energy tail and
large angular spread of the disrupted primary and secondary beams.

Previously, the extraction beam losses were evaluated \cite{nim} for the
IP beam parameters specified in the 2007 ILC Reference Design Report
(RDR) \cite{rdr}. Recently, the updated beam parameters have been
released in the 2013 ILC Technical Design Report (TDR) \cite{tdr}.
Therefore, it is important to verify the impact of the TDR
parameters on the extraction losses.

The study is performed for 500 GeV center-of-mass beam energy and the
extraction line optics \cite{exline} corresponding to the latest final
focus (FF) optics with $L^*=4.5$ m \cite{ff}, with and without detector
solenoid. The disrupted beam distribution at IP was generated using
Guinea-Pig code \cite{gpig}, and the beam losses were evaluated in tracking
simulations using DIMAD \cite{dimad}. The results using the TDR
parameters are compared with the results for the RDR ``nominal'' and
``low power'' (low-P) options as they have the parameters
closest to the TDR ones.

\vspace{-0mm}
\section{Beam parameters}

A beam collision, assuming the ILC parameters, will result in a strongly
disrupted distribution of the electron beam having a long low energy
tail and large angular spread. Trajectories of the low energy
electrons are then further amplified by strong deflections ($\propto\!1/E$)
in the extraction line magnets which can lead to beam losses. The
beam collision also creates a flux of beamstrahlung photons which can
contribute to the total power loss.

In this study, the beam losses are determined by tracking the electron
and photon beams from the IP to the extraction dump using DIMAD.
The particles are considered lost when their trajectories exceed the
size of the specified beam pipe aperture.
The initial disrupted distributions of electrons and photons after
collision were generated using Guinea-Pig code.

Tracking of $10^4-10^5$ electrons is typically sufficient for an
estimate of beam properties after IP and for calculation of
relatively high power loss in the extraction line collimators.
However, an accurate estimate of much lower losses in magnets, which are
mostly caused by low energy electrons, requires a higher statistics in
the low energy tail corresponding to $10^6-10^7$ particles in the full
beam. Tracking so many particles, however, takes much longer
time and requires rather large storage space for the data.
Fortunately, for this purpose it is not necessary to track the full beam
because only electrons with very low energy and large IP $x-y$ angles are
lost in the magnets. As a result, two types of beam data were used
in this study. The low statistics data, typically containing $4\cdot10^4$
electrons in the full beam, were used for evaluation of beam
distribution and calculation of high beam loss in the extraction
collimators. The high statistics beams with about $4\cdot10^6$ electrons
at IP were also generated, but only particles with energy below 70\% of
the nominal energy or $x$ or $y$ angles larger than 0.5 mrad at IP
were used in tracking for calculating beam losses in magnets. The mentioned
cuts reduce the beam population by a factor of 100.

The electron beam parameters for the TDR and the RDR nominal and
low-P options are summarized in Table \ref{tab:param}. Here, $N_e$ is the
number of electrons per bunch, $N_b$ the number of
bunches per pulse, $f_{RF}$ the repetition rate, $P$ the total beam power,
$\beta^*_{x,y}$ the beta functions at IP, $\sigma_z$ the bunch length,
$\gamma\epsilon_{x,y}$ the normalized emittances, $D_{x,y}$ the disruption
parameters, and $\delta_{BS}$ the fractional $rms$ energy loss due to
beamstrahlung.

\begin{table}[htb]
\vspace{-2mm}
\begin{center}
\caption{Design electron beam parameters at IP in the TDR and the RDR
nominal and low-P options.}
\vspace{1mm}
\small
\begin{tabular}{lccccccccccccc} \hline
& $E$ & $N_e$ & $N_b$ & $f_{RF}$ & $P$ & $\beta_x^*$ & $\beta_y^*$
& $\sigma_z$ & $\gamma\epsilon_x$ & $\gamma\epsilon_y$ & $D_x$ & $D_y$
& $\delta_{BS}$ \\
& GeV & $10^{10}$ & & Hz & MW & mm & mm & mm & $\mu m\cdot$rad
& $\mu m\cdot$rad & & & \% \\
\hline
TDR     & 250  & 2 & 1312 & 5 & 5.25 & 11 & 0.48 & 0.3 & 10 & 0.035
& 0.30  & 24.6 & 4.5 \\
Nominal & 250  & 2 & 2625 & 5 & 10.5 & 20 & 0.40 & 0.3 & 10 & 0.040
& 0.17  & 19.4 & 2.4 \\
Low-P   & 250  & 2 & 1320 & 5 & 5.29 & 11 & 0.20 & 0.2 & 10 & 0.036
& 0.21  & 26.1 & 5.5 \\
\hline
\end{tabular}
\normalsize
\label{tab:param}
\vspace{-6mm}
\end{center}
\end{table}

A larger value of $\delta_{BS}$ indicates a longer low energy tail
in the disrupted beam. Based on the $\delta_{BS}$ values in Table
\ref{tab:param} and taking into account that most of the electron losses occur
in the low energy tail, it is expected that beam losses with the TDR
beam parameters will be higher than in the RDR nominal option, but lower
than in the low-P option.

\vspace{-0mm}
\subsection{Disrupted electron distribution at IP with the TDR parameters}

\begin{figure}[htb]
\vspace{0mm}
\centering
\includegraphics*[height=75mm, angle=-90]{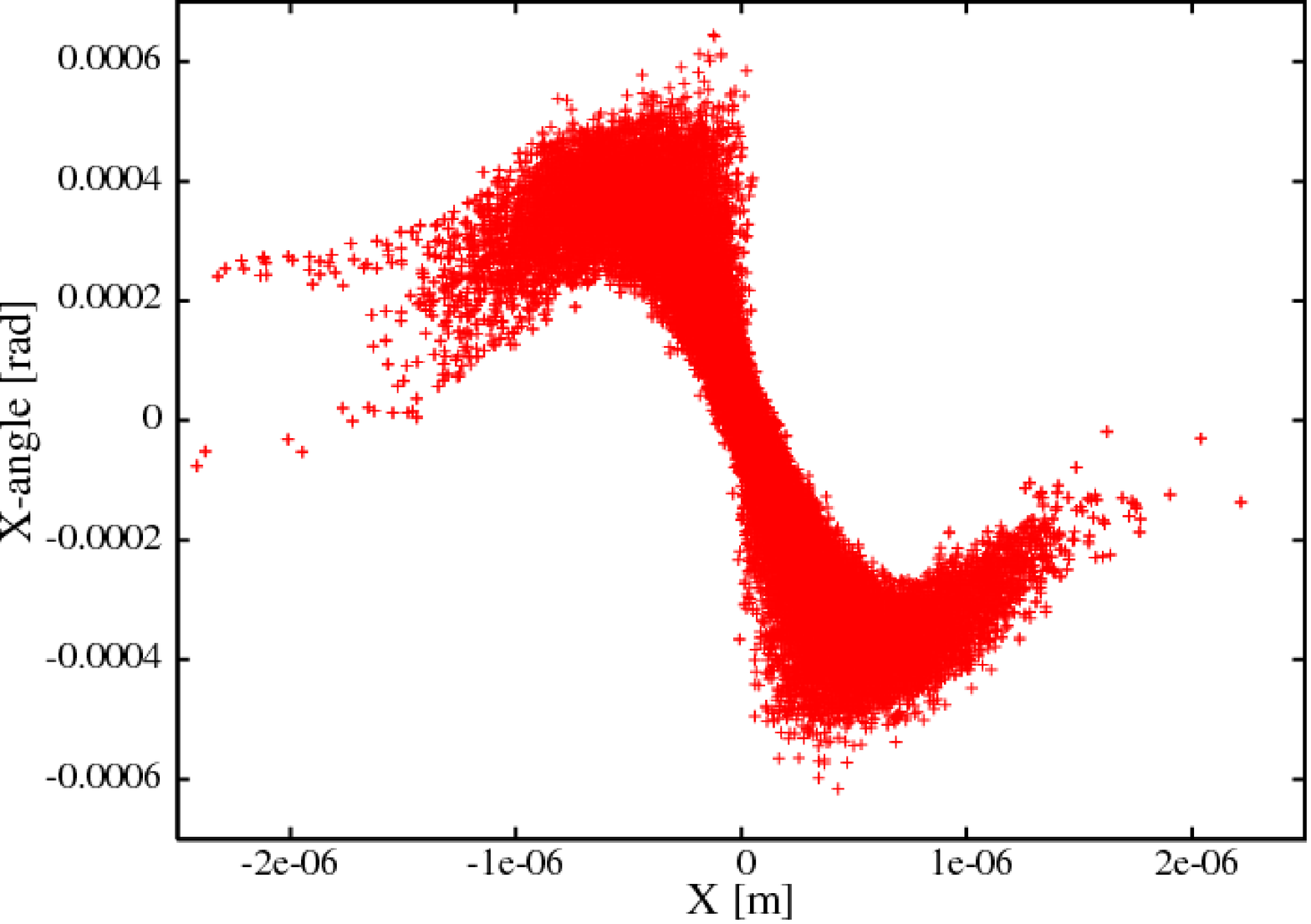}
\hspace{10mm}
\includegraphics*[height=75mm, angle=-90]{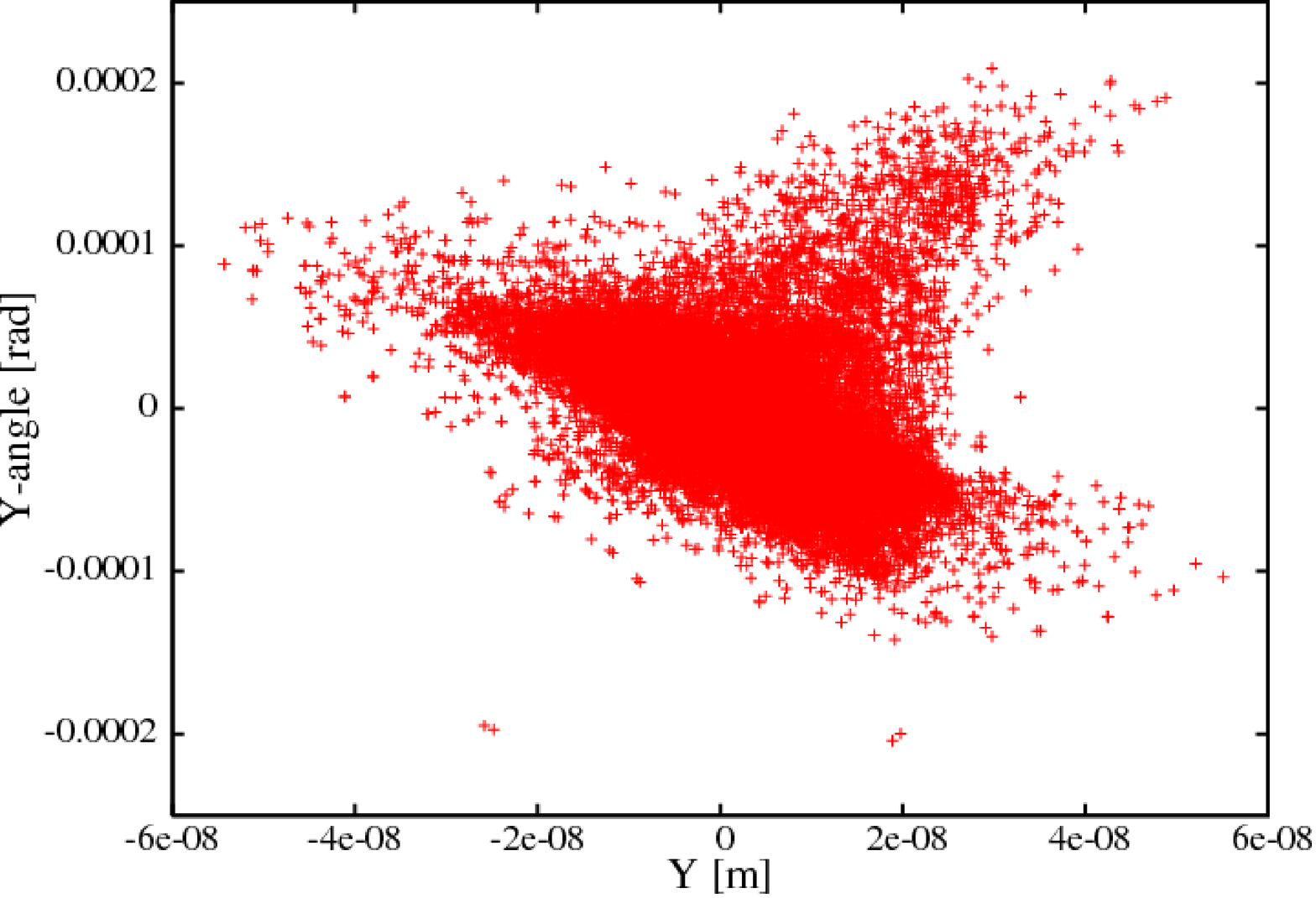}
\vspace{0mm}
\caption{Disrupted beam phase space at IP with the TDR parameters
for $4\cdot10^4$ electrons.}
\label{fig:phspace}
\vspace{0mm}
\end{figure}

\begin{figure}[htb]
\vspace{0mm}
\centering
\includegraphics*[width=75mm, angle=0]{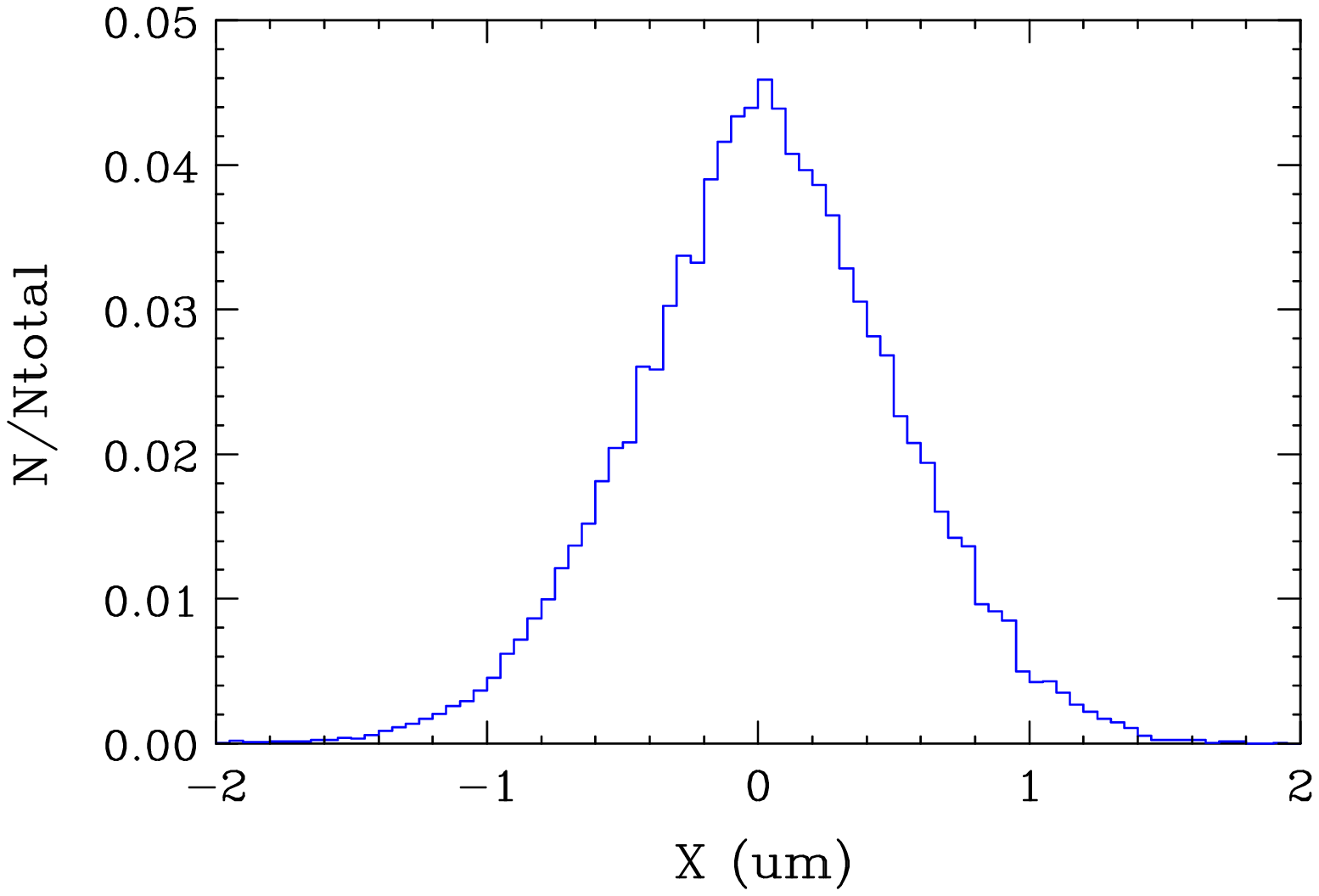}
\hspace{10mm}
\includegraphics*[width=75mm, angle=0]{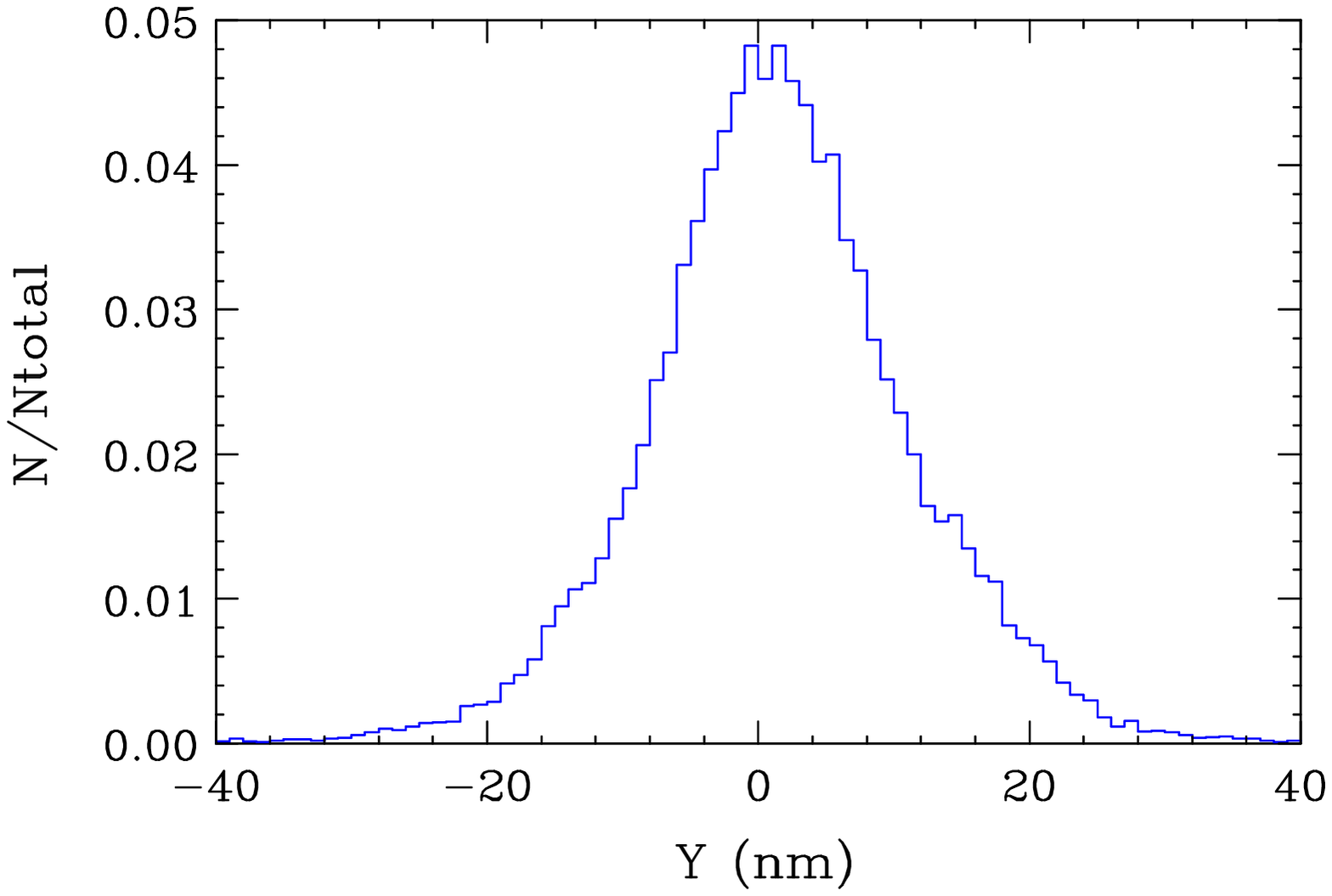}
\vspace{-2mm}
\caption{Disrupted horizontal ($\mu$m) and vertical (nm) distributions
at IP with the TDR parameters for $4\cdot10^4$ electrons.}
\label{fig:size}
\vspace{0mm}
\end{figure}

\begin{figure}[htb]
\vspace{0mm}
\centering
\includegraphics*[width=75mm, angle=0]{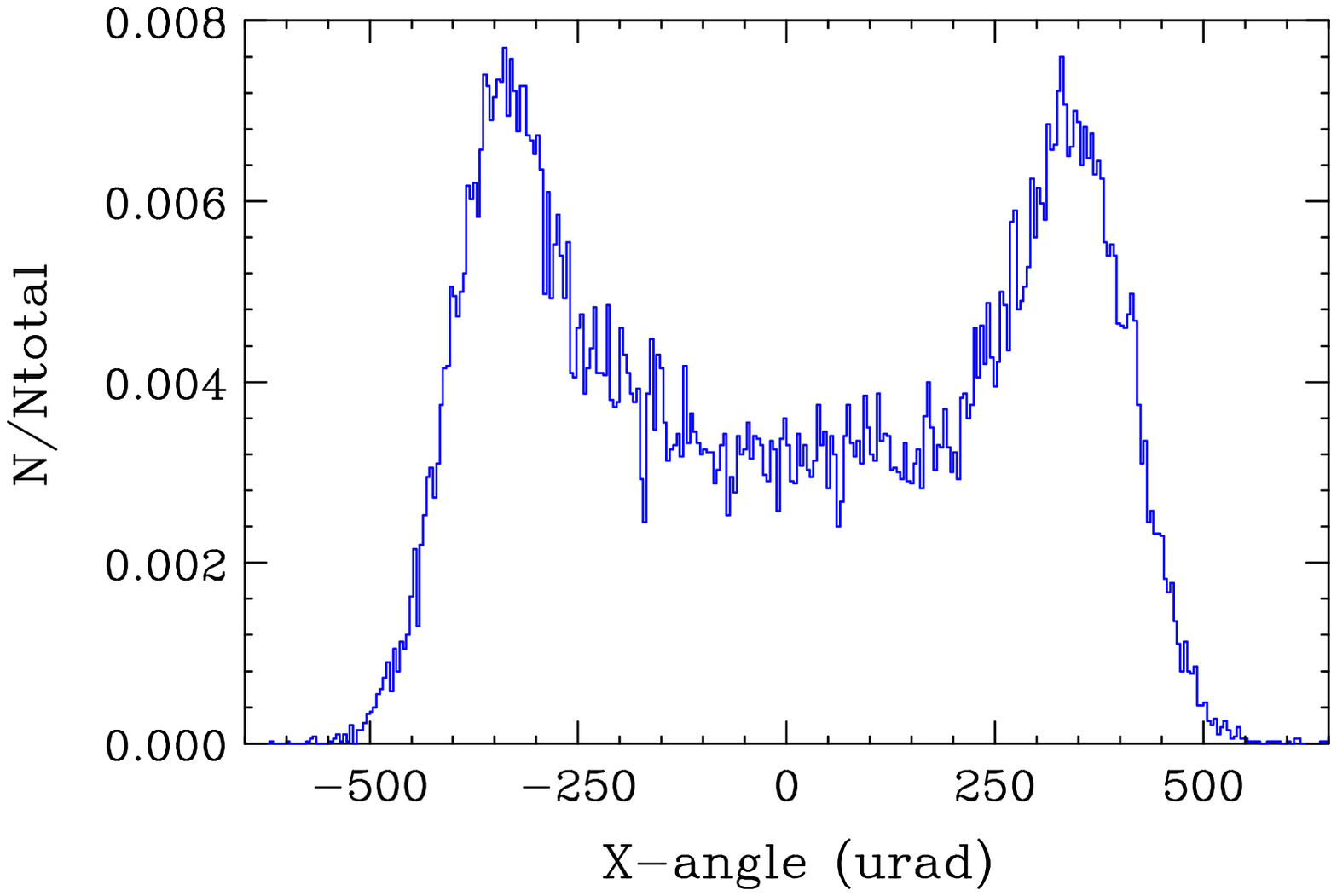}
\hspace{10mm}
\includegraphics*[width=75mm, angle=0]{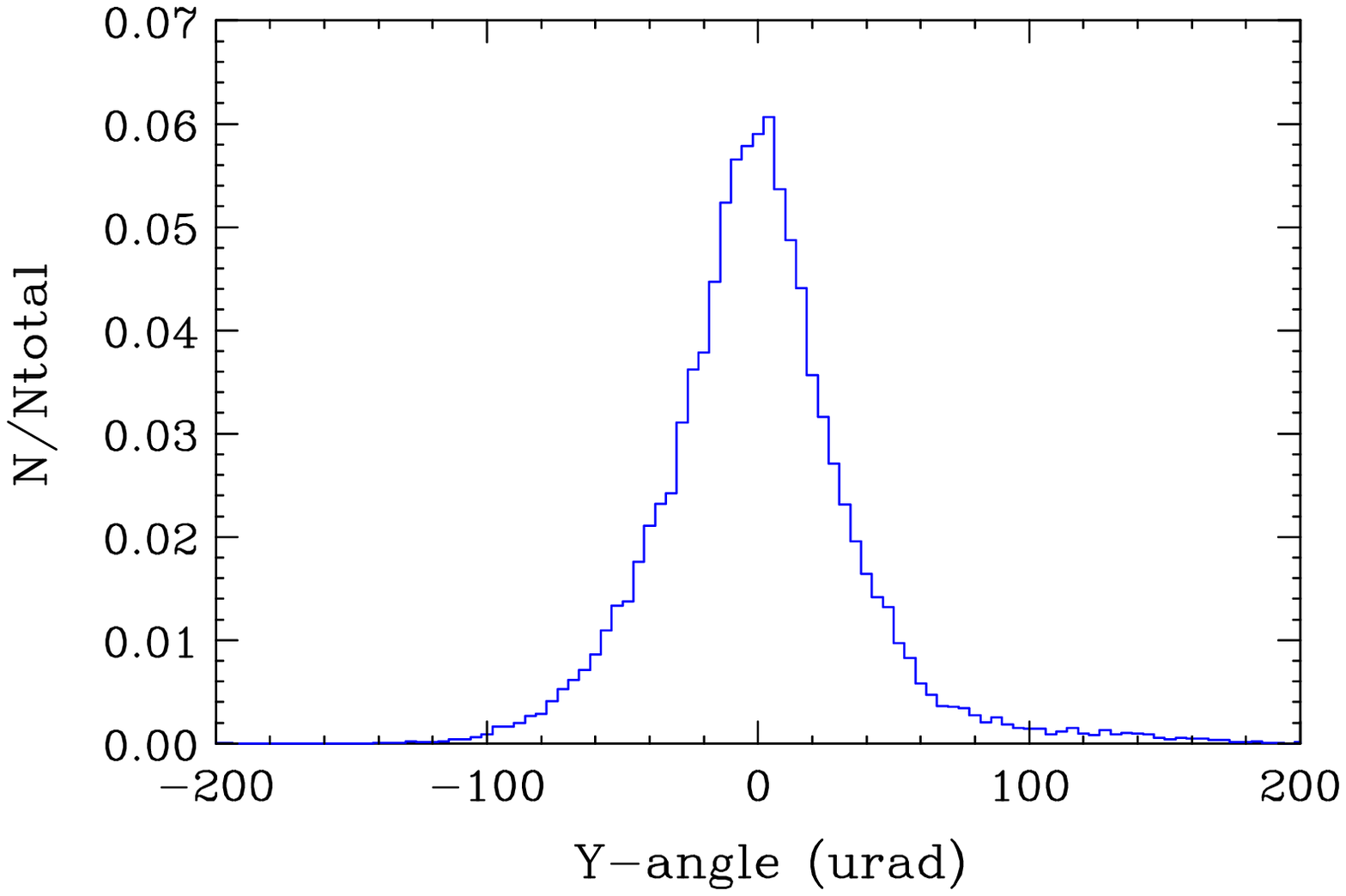}
\vspace{-2mm}
\caption{Disrupted angular distribution ($\mu$rad) at IP with the
TDR parameters for $4\cdot10^4$ electrons.}
\label{fig:angle}
\vspace{0mm}
\end{figure}

The impact of beam disruption on electron beam size
with the TDR parameters can be seen in Table \ref{tab:size}.
The values for the undisrupted beam correspond to the parameters in
Table \ref{tab:param}, and the values for disrupted beam are computed
from the electron distribution generated by Guinea-Pig.
One can see that the disruption effect significantly increases the
transverse angular spread and the beam emittance.
Fig. \ref{fig:phspace}-\ref{fig:angle} show the $x$ and $y$
phase space, beam size and angular distributions for the disrupted beam
with the TDR parameters.
Note the sloped horizontal phase space in Fig. \ref{fig:phspace} and
double peak horizontal angular spread in Fig. \ref{fig:angle}
characteristic for flat beam collisions.

\begin{table}[htb]
\vspace{-2mm}
\begin{center}
\caption{RMS beam size, beam angular divergence and emittance at IP
with and without disruption for $4\cdot10^4$ electrons.}
\vspace{1mm}
\begin{tabular}{lcccccc} \hline
IP beam & $\sigma_x^*$ & $\sigma_y^*$ & $\sigma_{x^\prime}^*$ &
$\sigma_{y^\prime}^*$ & $\gamma\epsilon_x$ & $\gamma\epsilon_y$ \\
parameters & nm & nm & $\mu$rad & $\mu$rad &
$\mu m\cdot$rad & $\mu m\cdot$rad \\
\hline
undisrupted & 474 & 5.9 &  43 & 12.2 & 10 & 0.035 \\
disrupted   & 493 & 9.9 & 284 & 36.2 & 36 & 0.164 \\
\hline
\end{tabular}
\normalsize
\label{tab:size}
\vspace{-6mm}
\end{center}
\end{table}

\begin{figure}[htb]
\vspace{0mm}
\centering
\includegraphics[height=100mm, angle=-90]{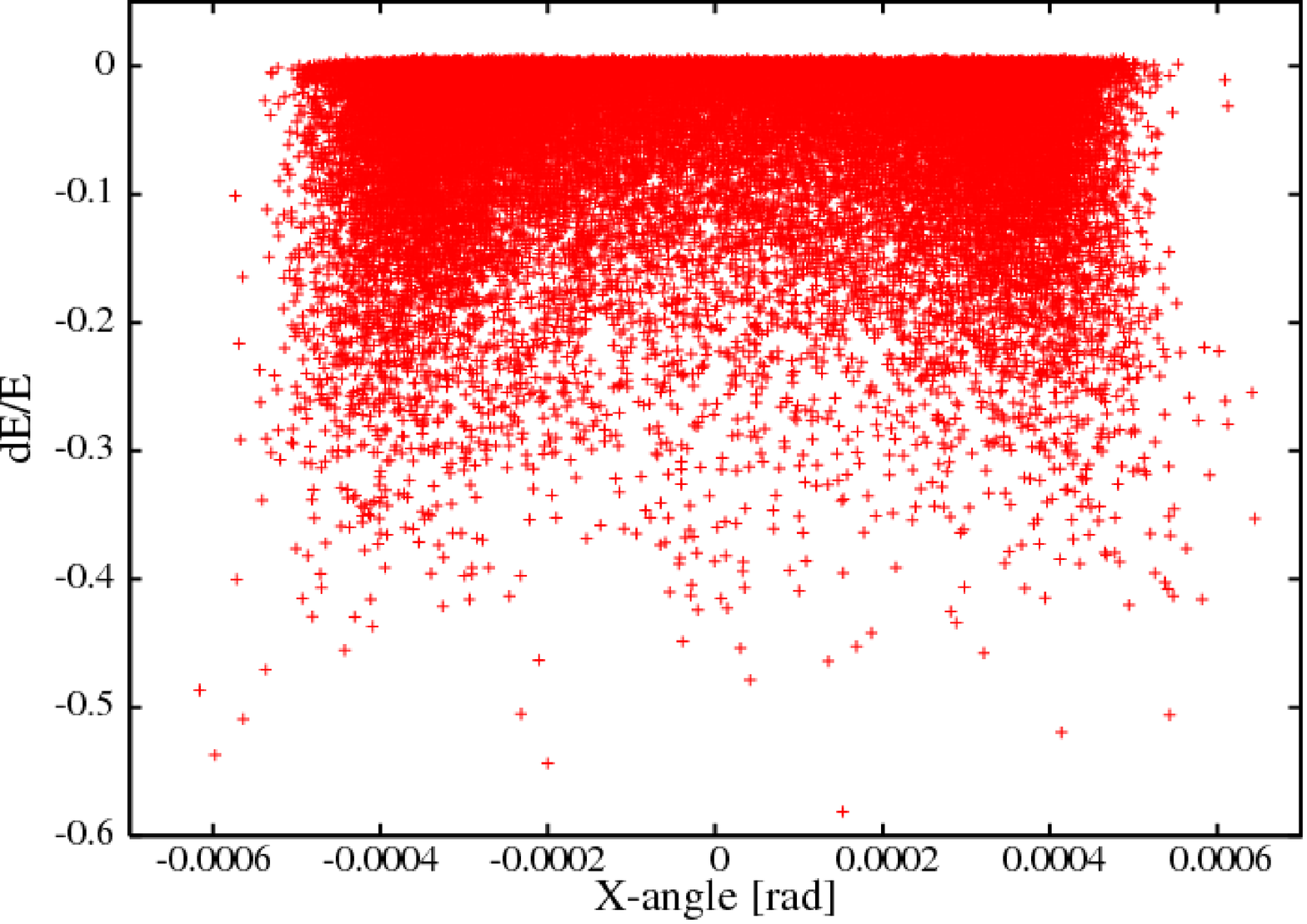}
\vspace{0mm}
\caption{Disrupted $\Delta E/E$ versus horizontal angle at IP
with the TDR parameters for $4\cdot10^4$ electrons.}
\label{fig:dEpx}
\vspace{0mm}
\end{figure}

\begin{figure}[htb]
\vspace{0mm}
\centering
\includegraphics[width=100mm, angle=0]{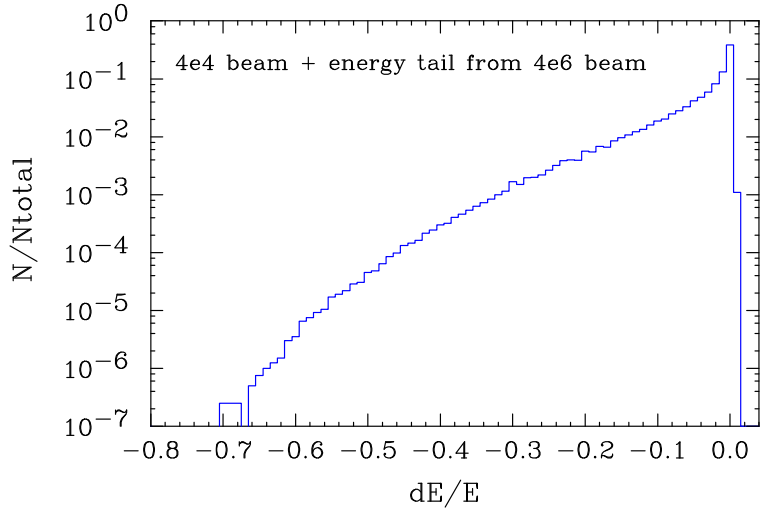}
\vspace{-2mm}
\caption{Disrupted energy distribution at IP with the TDR parameters
for $4\cdot10^6$ electrons.}
\label{fig:dE}
\vspace{0mm}
\end{figure}

Fig. \ref{fig:dEpx}, \ref{fig:dE} show the distribution of relative
electron energy offsets $\Delta E/E$ in the disrupted beam with the TDR
parameters. Electrons in the low energy tail will experience
strong deflections ($\propto 1/E$) in the extraction magnets leading
to large trajectories and potential loss on beam chamber.
The low electron energy and large IP angles are the main
sources for the electron loss in the extraction magnets.

\vspace{-0mm}
\subsection{Beamstrahlung photon distribution at IP with the TDR parameters}

Bending of the electron orbits in collision due to beam-beam forces
results in radiation creating a flux of beamstrahlung
photons traveling in the same direction with the primary electrons.
The photons are not affected by the magnetic field of the extraction
magnets, therefore their trajectories are determined strictly by
the IP angles. Beam power in the beamstrahlung photon beam is not
negligible, therefore the extraction line must provide
sufficient aperture increasing with distance. The current design
has the magnet aperture accommodating photon angles up to
$\pm0.75$ mrad.

\begin{figure}[htb]
\vspace{0mm}
\centering
\includegraphics*[width=75mm, angle=0]{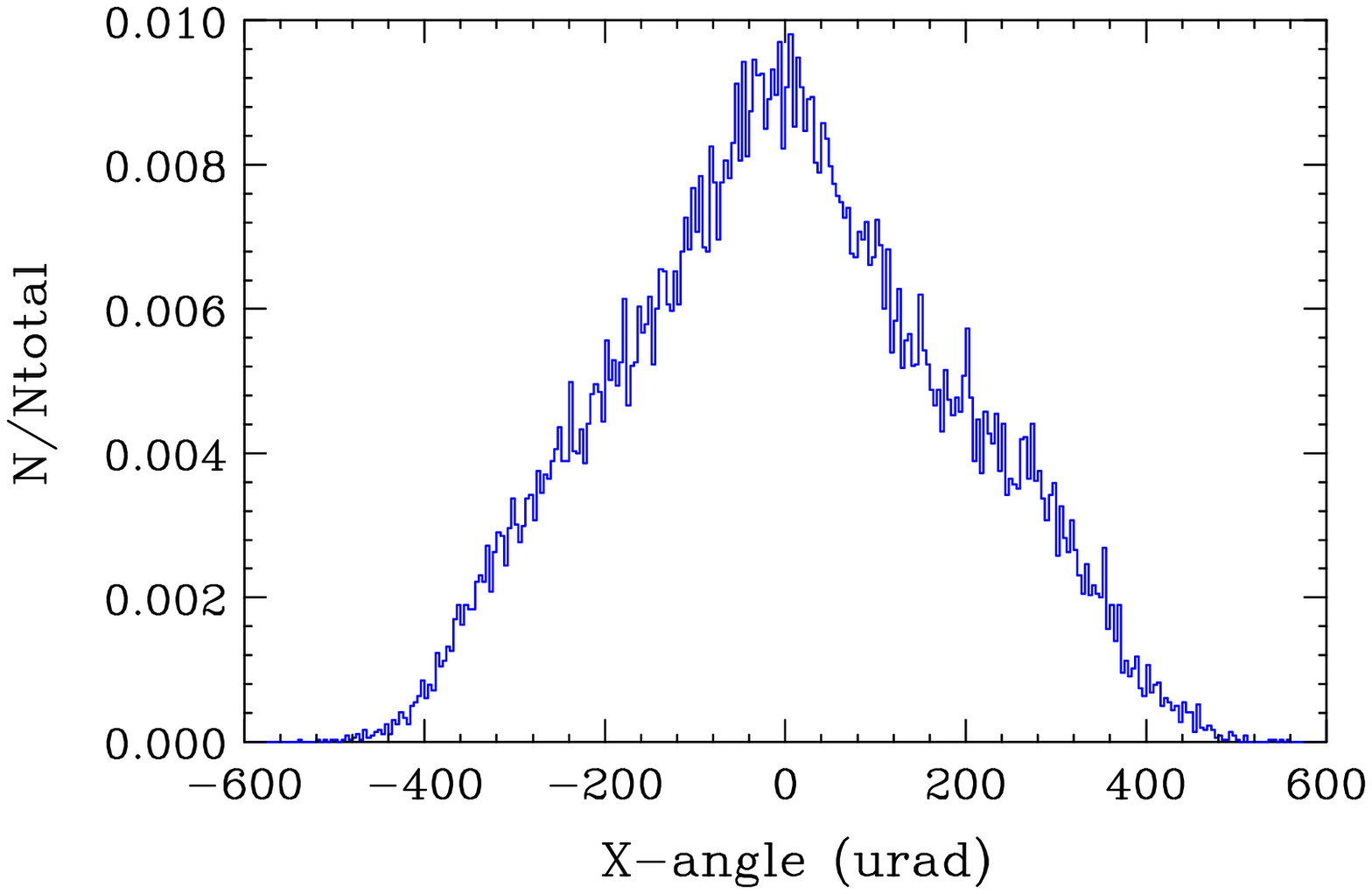}
\hspace{10mm}
\includegraphics*[width=75mm, angle=0]{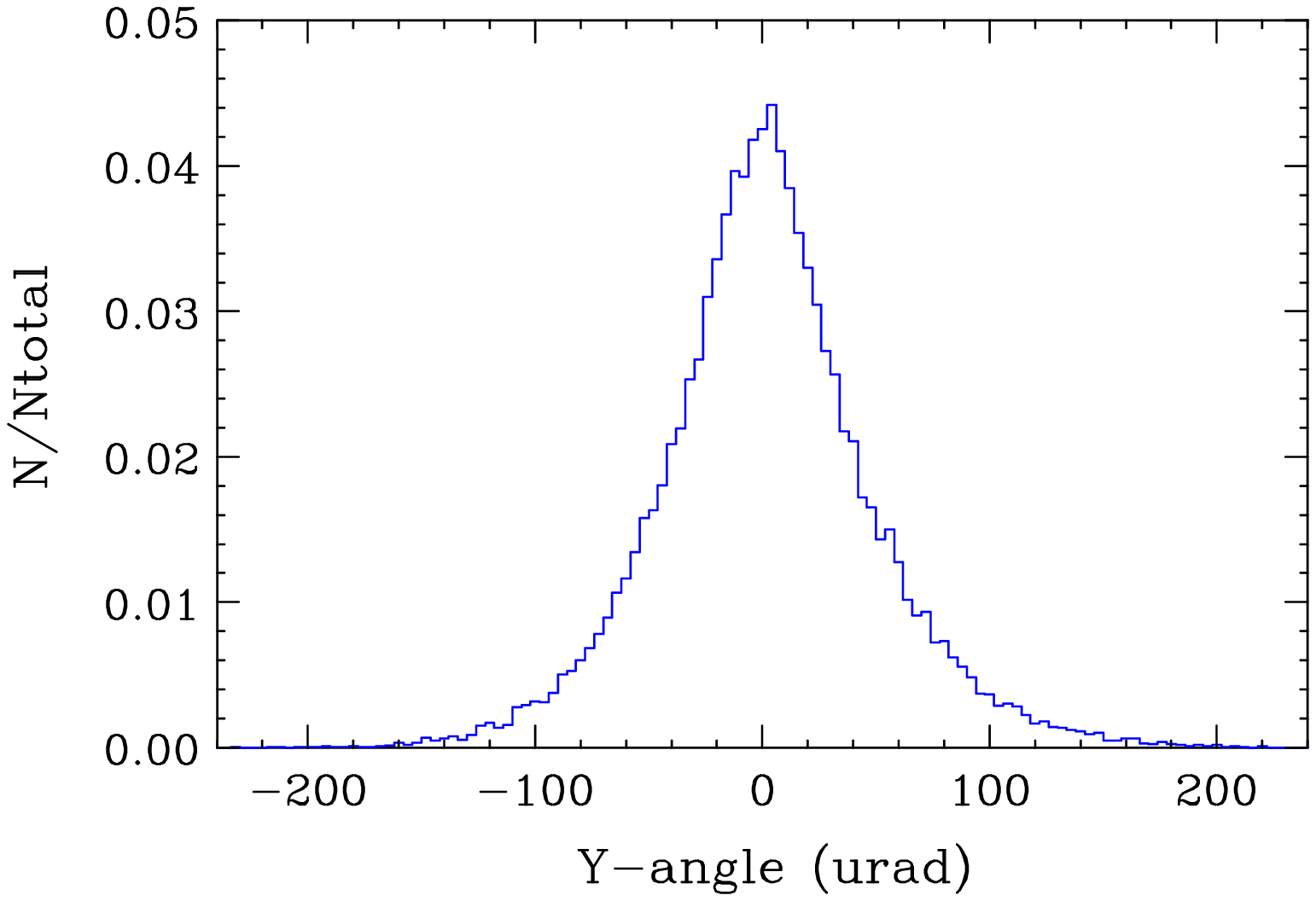}
\vspace{-2mm}
\caption{Beamstrahlung angular distribution ($\mu$rad) at IP with the
TDR parameters for $3.6\cdot10^4$ photons.}
\label{fig:photons}
\vspace{0mm}
\end{figure}

Table ~\ref{tab:photons} shows $rms$ and maximum
photon angles with the TDR parameters for $3.6\cdot10^4$ photons
generated by Guinea-Pig. Note that this angular spread is well
within the extraction magnet aperture, although larger angles can
occur if a higher statistics is generated or other mechanisms are
considered (e.g. non-ideal collisions). The beamstrahlung
angular $x$ and $y$ distributions at IP with the TDR parameters
are shown in Fig. \ref{fig:photons}.

\begin{table}[htb]
\vspace{-2mm}
\begin{center}
\caption{RMS divergence and maximum angles at IP for beam of
$3.6\cdot10^4$ beamstrahlung photons.}
\vspace{1mm}
\begin{tabular}{cccc} \hline
$\sigma_{x^\prime}^*$ & $\sigma_{y^\prime}^*$ &
$x^\prime_{max}$ & $y^\prime_{max}$ \\
\hline
184 $\mu$rad & 47 $\mu$rad & 559 $\mu$rad & 308 $\mu$rad \\
\hline
\end{tabular}
\normalsize
\label{tab:photons}
\vspace{-6mm}
\end{center}
\end{table}

\vspace{-0mm}
\section{Extraction line optics}

Detailed description of the extraction line optics can be found in
\cite{exline}. The design accommodates a range of $L^*$ in the FF from
3.5 m to 4.5 m. In this study, the FF optics with $L^*=4.5$ m \cite{ff}
is used. The corresponding extraction optics has a free space
$L_{ext}^*=6.3$ m between IP and the nearest extraction quadrupole.

\begin{figure}[htb]
\vspace{0mm}
\centering
\includegraphics*[height=140mm, angle=-90]{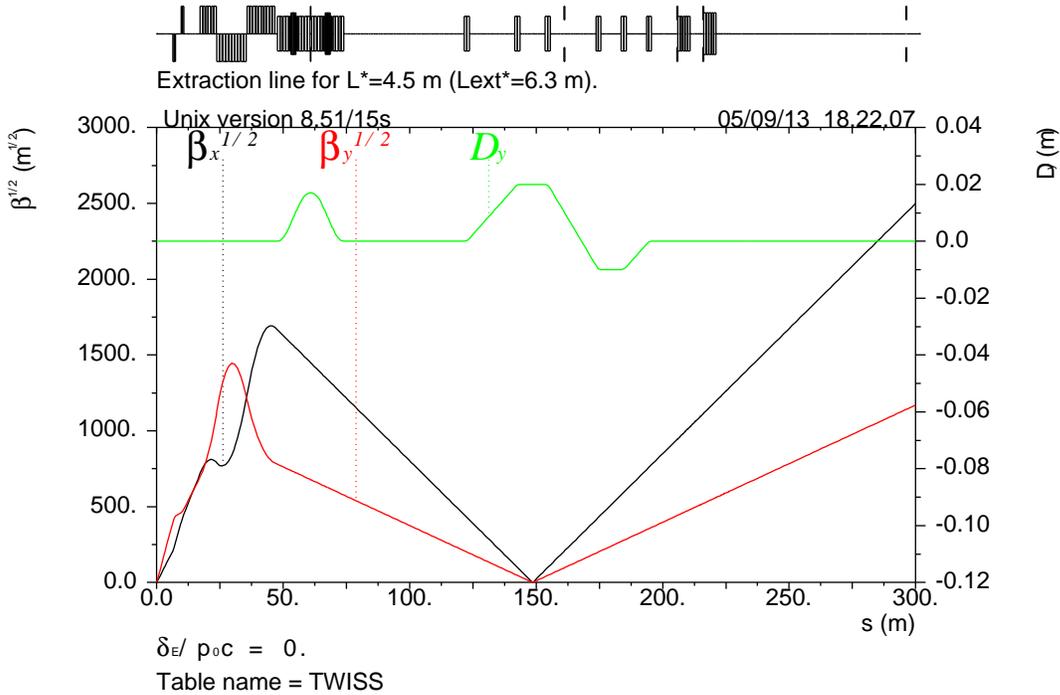}
\vspace{0mm}
\caption{Extraction line layout (top) and optics functions with the
TDR parameters.}
\label{fig:optics}
\vspace{0mm}
\end{figure}

The extraction line layout and optics functions for the TDR parameters
are shown in Fig. \ref{fig:optics}. The quadrupole focusing in the
beginning of the line is designed to provide a secondary focal point
with an optimal transformation for polarization measurement, and large
chromatic and geometric acceptance for the electron and photon beams.
Downstream of the quadrupoles, the optics includes six vertical dipoles
providing conditions for measurements of beam energy,
polarization and luminosity \cite{diag}. After the last dipole, there
are 5 horizontal and 5 vertical fast sweeping kickers
which function is to increase the effective beam area at the dump for
protecting the dump window from high beam power density and preventing
water boiling in the dump.
The extraction collimation system includes two protection collimators within
the dipole region and three collimators within the final 100 m before
the dump. The latter protect the sweeping kickers and limit beam area to
within 15 cm radius at the dump window.

As noted earlier, the disruption significantly changes the electron beam
distribution. Consequently, this modifies the IP optics
functions which can be computed from the statistical electron distribution
generated by Guinea-Pig. The $\beta$ and $\alpha$ functions at IP before
and after disruption are shown in Table \ref{tab:ipbeta}.

\begin{table}[htb]
\vspace{-2mm}
\begin{center}
\caption{Optics functions at IP for disrupted and undisrupted electron
beam at IP with the TDR parameters.}
\vspace{1mm}
\begin{tabular}{lcccc} \hline
IP beam & $\beta_x^*$ & $\alpha_x^*$ & $\beta_y^*$ & $\alpha_y^*$ \\
\hline
undisrupted & 11 mm   & 0     & 0.48 mm  & 0 \\
disrupted   & 3.29 mm & 1.609 & 0.294 mm & 0.386 \\
\hline
\end{tabular}
\normalsize
\label{tab:ipbeta}
\vspace{-6mm}
\end{center}
\end{table}

Finally, a 5 T detector solenoid based on a silicon detector design (SiD)
\cite{SiD} is included in the tracking study. The orbit distortions
due to the solenoid are compensated using dipole corrector windings
in the superconducting (SC) extraction quadrupoles \cite{nim}. Additionally,
a detector integrated dipole coil (anti-DiD) built on top of the
solenoid is included as it helps minimizing the detector background
\cite{DiD}.

\vspace{-0mm}
\section{Disrupted electron beam losses for the TDR
parameters without detector solenoid}

\begin{figure}[htb]
\vspace{0mm}
\centering
\includegraphics*[height=100mm, angle=-90]
{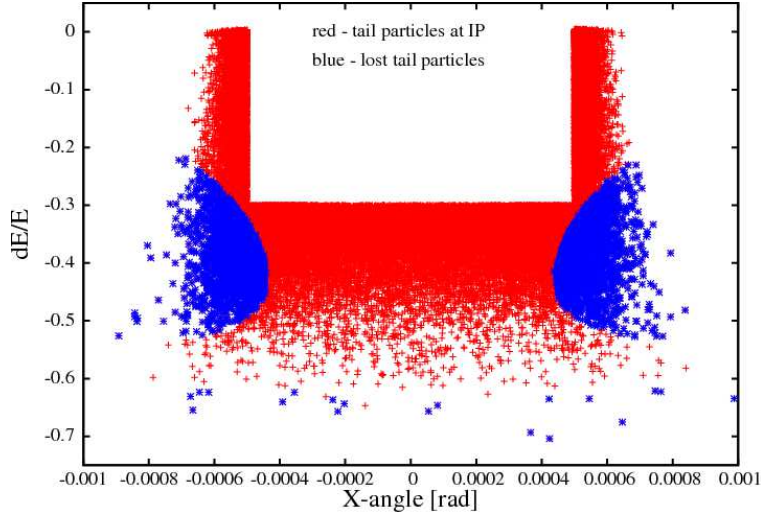}
\vspace{0mm}
\caption{Initial energy and angular distribution of tail electrons with
$E<0.7\,E_0$ or $x$ or $y$ angles $>0.5$ mrad at IP (red), and
distribution of lost electrons in the extraction line (blue)
with the TDR parameters without detector solenoid.}
\label{fig:tailnosol}
\vspace{0mm}
\end{figure}

\begin{figure}[htb]
\vspace{0mm}
\centering
\includegraphics*[width=100mm, angle=0]
{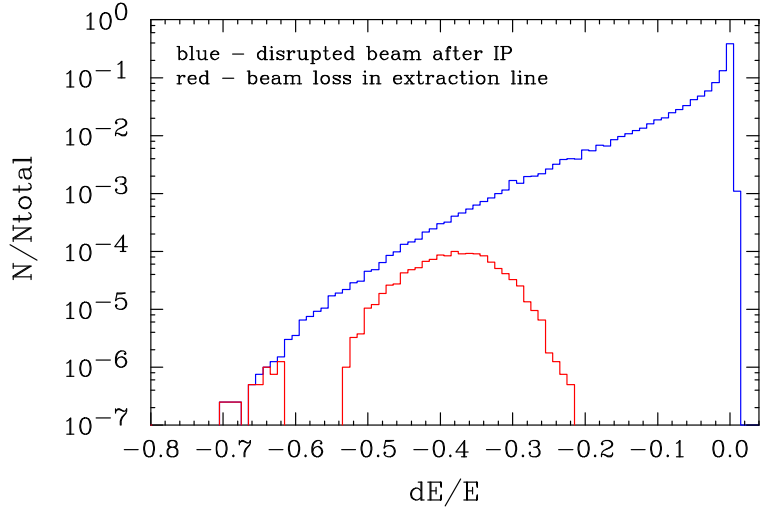}
\vspace{-2mm}
\caption{Initial disrupted electron energy distribution (blue), and
distribution of lost electrons in the extraction line (red) with the
TDR parameters without detector solenoid.}
\label{fig:dEnosol}
\vspace{0mm}
\end{figure}

The extraction beam losses occur when electrons have either low energy
or large $x$ or $y$ angles at IP. The tracking simulations show that in
order to determine the complete losses, it is sufficient to track only
the beam tail where energy is below 70\% of the nominal energy or IP
$x$ or $y$ angles are larger than 0.5 mrad (this accounts for $\approx1.3\%$
electrons of the total beam with the TDR parameters).
Only 52289 particles out of the high statistics beam of $4\cdot10^6$ particles
satisfy the mentioned criteria. The resulting beam losses without the detector
solenoid amounted to 5062 electrons.
Fig. \ref{fig:tailnosol} shows the initial distribution of $\Delta E/E$ 
and horizontal angles at IP in the tracked beam tail, and distribution
of electrons lost in the extraction line. One can see that, indeed, 
the lost electrons belong
to the energy and angular tail as specified above. Note also the missing
beam core in Fig. \ref{fig:tailnosol} which was not tracked for speeding
up the calculation.
Fig. \ref{fig:dEnosol} shows the initial and lost electron energy
distribution for the full beam, without the solenoid.

\begin{figure}[htb]
\vspace{0mm}
\centering
\includegraphics*[width=100mm, angle=0]
{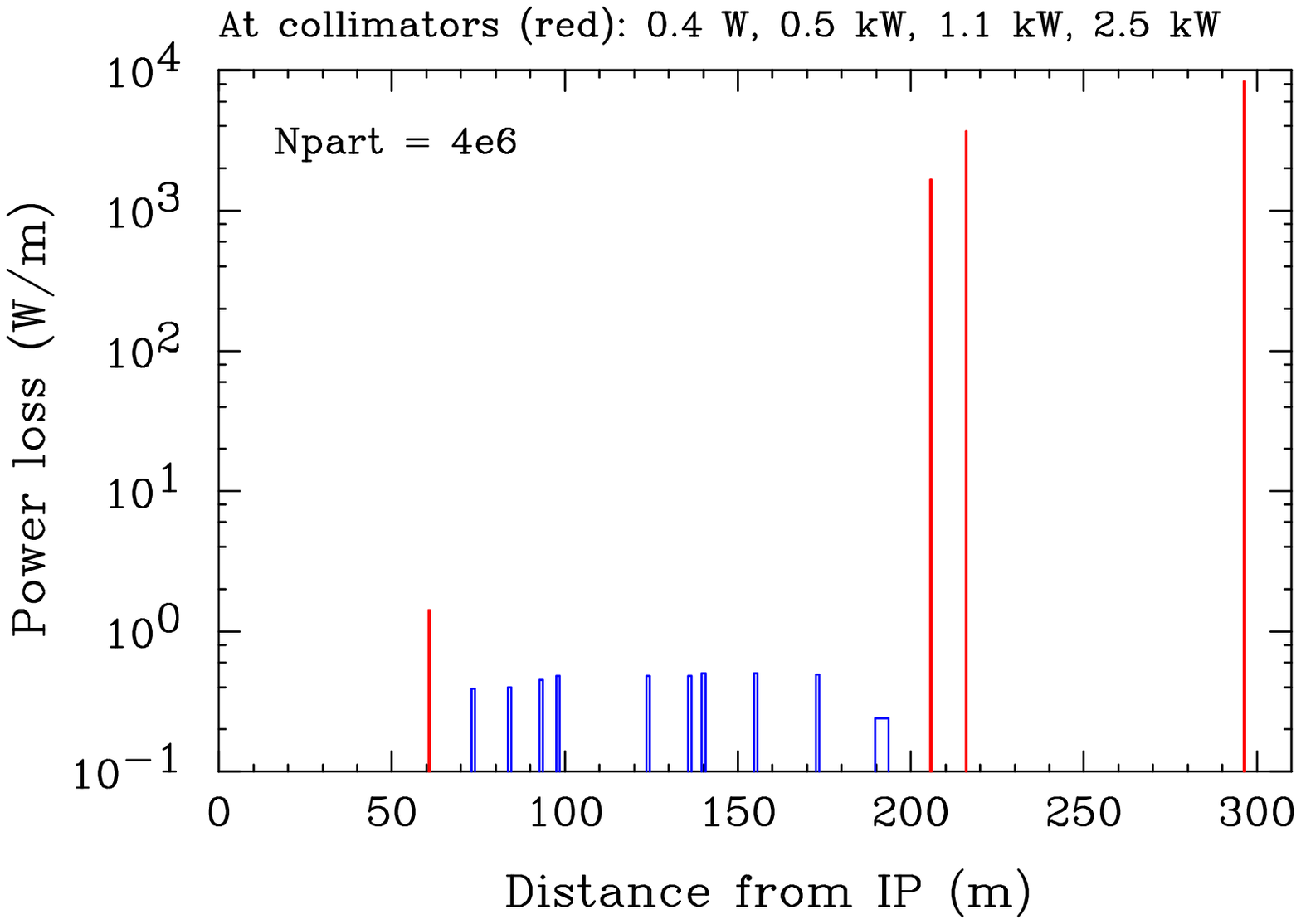}
\vspace{0mm}
\caption{Power losses of the disrupted electron beam in the extraction
line magnets and diagnostics (blue), and collimators (red) with the
TDR parameters without detector solenoid.}
\label{fig:powernosol}
\vspace{0mm}
\end{figure}

Based on the electron loss data, one can obtain power losses in the
extraction magnets, diagnostics and collimators as shown in Fig.
\ref{fig:powernosol}.
Most power losses occur in the collimators designed to protect
the magnets and diagnostics as well as to control the beam spread at the dump.
In this case, no loss was observed at the SC quadrupoles, and the losses
in warm magnets are below 0.5 W/m which is acceptable. The
maximum power loss in a collimator is 2.5 kW which is acceptable.

\vspace{-0mm}
\section{Disrupted electron beam losses for the TDR
parameters with SiD detector solenoid}

Similarly, the electron beam losses with the TDR parameters were
calculated including the SiD detector solenoid and the anti-DiD dipole
field. In this case, due to the crossing angle, the solenoid generates
vertical orbit and dispersion as well as coupling to the beam.
It is also assumed that the incoming orbit at IP has a non-zero vertical
angle due to the effect of the upstream part of the solenoid.
A conservatively large angle value of $y^\prime=100$ $\mu$rad is used
in the tracking.

\begin{figure}[htb]
\vspace{0mm}
\centering
\includegraphics*[height=100mm, angle=-90]
{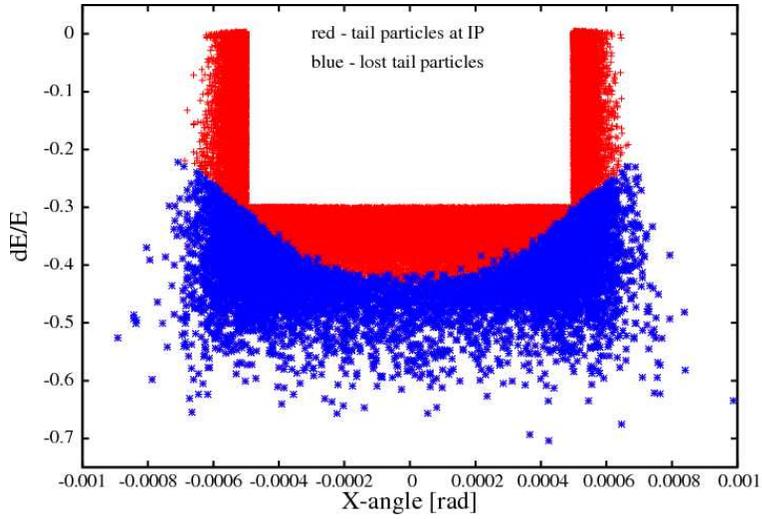}
\vspace{0mm}
\caption{Initial energy and angular distribution of tail electrons with
$E<0.7\,E_0$ or $x$ or $y$ angles $>0.5$ mrad at IP (red), and
distribution of lost electrons in the extraction line (blue)
with the TDR parameters and SiD detector solenoid.}
\label{fig:tailsol}
\vspace{0mm}
\end{figure}

\begin{figure}[htb]
\vspace{0mm}
\centering
\includegraphics*[width=100mm, angle=0]
{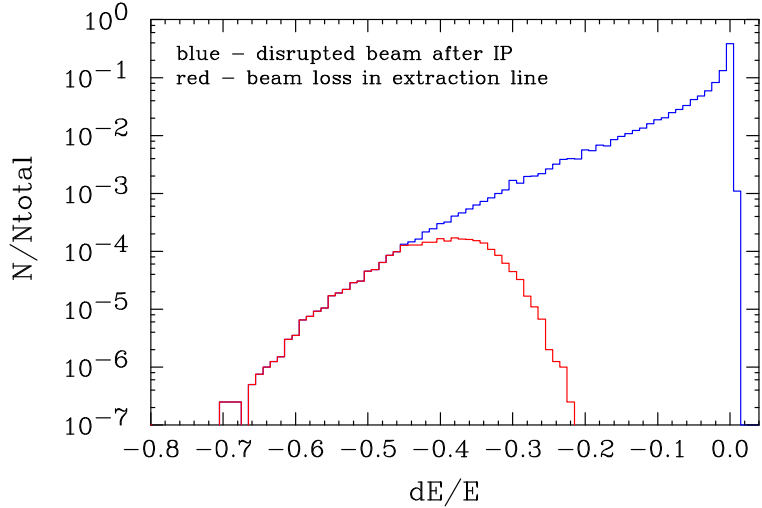}
\vspace{-2mm}
\caption{Initial disrupted electron energy distribution (blue), and
distribution of lost electrons in the extraction line (red) with the
TDR parameters and SiD detector solenoid.}
\label{fig:dEsol}
\vspace{0mm}
\end{figure}

The first order orbit from the solenoid is locally corrected by the
four dipole corrector windings included in the SC quadrupoles. However,
the remaining residual chromatic and coupling distortion can increase the
beam loss. The beam tail distribution at IP, as in the case
without solenoid, was used in tracking with the SiD solenoid. The resulting
energy and angular distributions of the initial and lost electrons are
shown in Fig. \ref{fig:tailsol} and \ref{fig:dEsol}.
One can see that most electrons with low energy, regardless of initial
$x-y$ angles, are lost. This effect is likely due to residual dispersion from
the solenoid. The total electron losses are increased by a factor of 2 compared
to the case without solenoid. These additional losses can be reduced if the
incoming vertical angle at IP is compensated.

\begin{figure}[htb]
\vspace{0mm}
\centering
\includegraphics*[width=100mm, angle=0]
{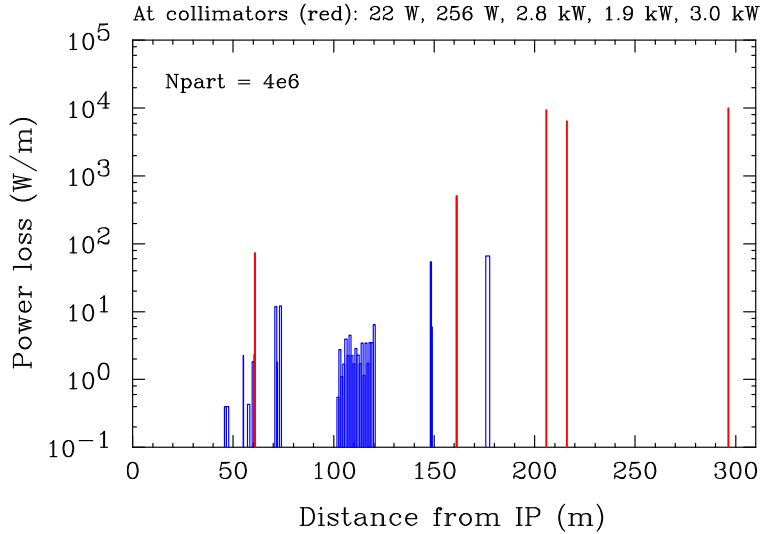}
\vspace{-2mm}
\caption{Power losses of the disrupted electron beam in the extraction
line magnets and diagnostics (blue), and collimators (red) with the TDR
parameters and SiD detector solenoid.}
\label{fig:powersol}
\vspace{0mm}
\end{figure}

The electron power losses with the TDR parameters and the SiD solenoid
are shown in Fig. \ref{fig:powersol}.
Most of the losses occur in the protection collimators due to their tighter
apertures. No loss was observed in the SC quadrupoles, and losses
in the warm magnets are below 12 W/m which should be acceptable.
The maximum power loss in a collimator is 3 kW which is acceptable.

\vspace{-0mm}
\section{Electron beam power losses in the RDR nominal and low-P options
with SiD detector solenoid}

For comparison, the electron beam power losses
were also calculated for the RDR nominal and low-P options, including
the SiD detector solenoid and the anti-DiD dipole coil field. In the nominal
option, the tracking was performed using a beam tail extracted
from $3.5\cdot10^7$ beam data. In case of the low-P parameters,
where disruption is much stronger, the tracking was done with a full
beam containing $7\cdot10^4$ particles. The same incoming 
vertical angle of 100 $\mu$rad at IP was used as in the TDR tracking.

\begin{figure}[htb]
\vspace{0mm}
\centering
\includegraphics*[width=100mm, angle=0]
{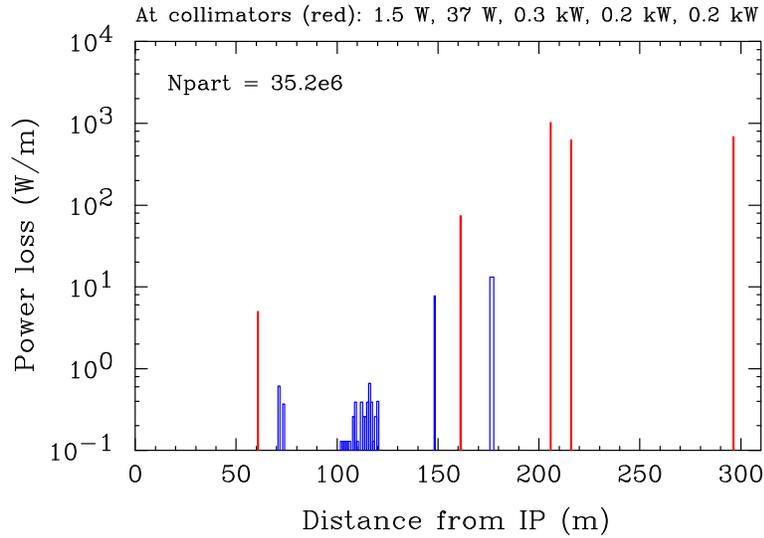}
\vspace{-2mm}
\caption{Power losses of the disrupted electron beam in the extraction
line magnets and diagnostics (blue), and collimators (red) with the
SiD detector solenoid and RDR nominal parameter option.}
\label{fig:nominal}
\vspace{0mm}
\end{figure}

\begin{figure}[htb]
\vspace{0mm}
\centering
\includegraphics*[width=100mm, angle=0]
{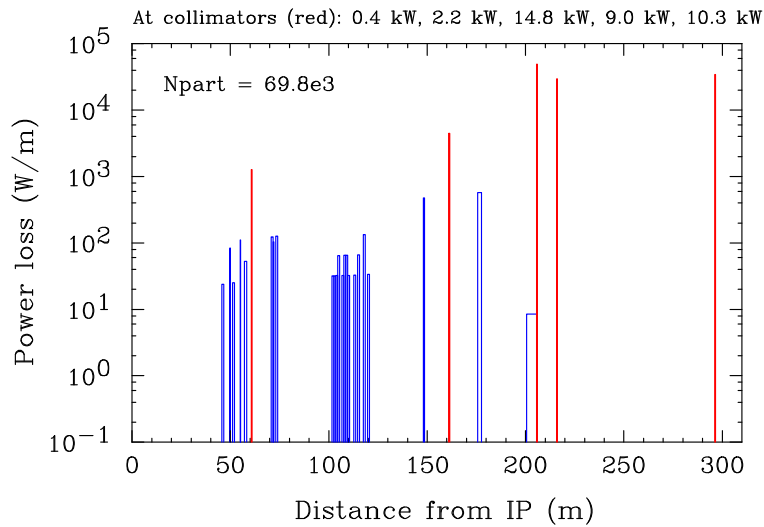}
\vspace{-2mm}
\caption{Power losses of the disrupted electron beam in the extraction
line magnets and diagnostics (blue), and collimators (red) with the
SiD detector solenoid and RDR low-P parameter option.}
\label{fig:lowP}
\vspace{0mm}
\end{figure}

Power losses in the RDR nominal and low-P options are shown in Fig.
\ref{fig:nominal} and \ref{fig:lowP}. The losses in the nominal option
are lower compared to the TDR case due to smaller disruption resulting
in a shorter low energy tail.

The losses in the low-P option are significantly higher compared to
both the TDR and RDR nominal options due to higher
disruption. Although there is no loss in the SC quadrupoles, the
power losses in warm magnets are up to 130 W/m, and there are
rather high losses in the diagnostic detectors (at $S\approx150$ m
and 175 m) which may be a concern. The maximum power loss at
a collimator is 14.8 kW which can be acceptable with a proper
design.

\vspace{-0mm}
\section{Beamstrahlung photon power losses with the TDR parameters}

Since the photon trajectories are not affected by magnetic field, their
losses are determined only by their $x-y$ angles at IP and the extraction
aperture. By design, the extraction aperture accepts the IP photon angles
up to $\pm0.75$ mrad within the magnets and diagnostics ($S=0$ to 200 m),
and $\pm0.5$ mrad in the dump collimators ($S=200$ to 300 m).
Tracking was performed for $3.6\cdot10^4$ beamstrahlung photons
corresponding to $2\cdot10^4$ primary electrons. In agreement with
the photon angular distribution in Fig. \ref{fig:photons}, where maximum
angle is below 0.75 mrad, there were no photon losses in magnets and
diagnostics. A rather small loss of 100 W occurred at two protection
collimators about 90 m upstream of the dump.

\begin{figure}[htb]
\vspace{0mm}
\centering
\includegraphics*[width=100mm, angle=0]
{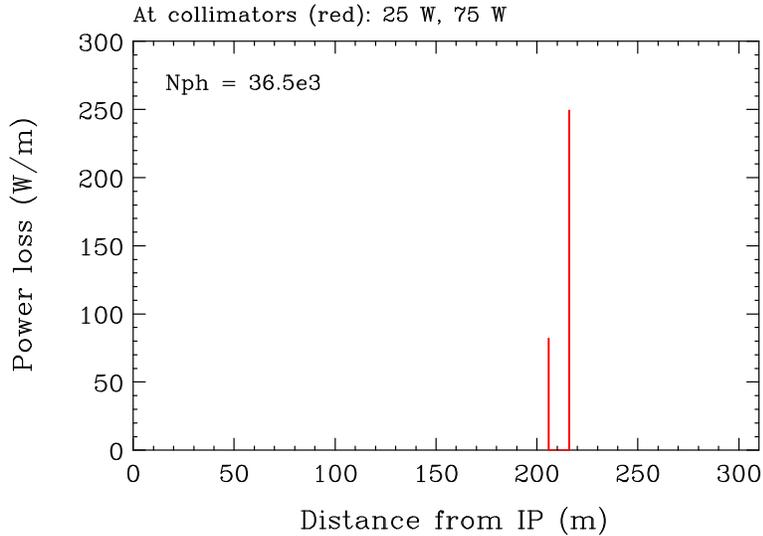}
\vspace{0mm}
\caption{Power losses of beamstrahlung photons in the extraction line
collimators with the TDR parameters. No losses occurred in magnets and
diagnostics for $3.6\cdot10^4$ tracked photons.}
\label{fig:powerphoton}
\vspace{0mm}
\end{figure}

\vspace{-0mm}
\section{Summary of extraction losses and conclusions}

Table \ref{tab:summary} summarizes electron power losses in magnets,
diagnostic detectors and collimators for the TDR parameters and the RDR
nominal and low-P options with the SiD detector solenoid.
One can see that the power losses in extraction magnets with the TDR
parameters are an order of magnitude higher than in the RDR nominal
option, but an order of magnitude lower than in the RDR low-P option.
Overall, the electron beam losses in the extraction magnets and
collimators with the TDR parameters appear to be acceptable. Losses in the
diagnostic detectors may need an expert opinion to evaluate the impact
on the background signal. The beamstrahlung photon losses with the
TDR parameters are rather small.

\begin{table}[htb]
\vspace{0mm}
\begin{center}
\caption{Summary of disrupted electron beam loss in the extraction line
with the SiD detector solenoid for the TDR parameters and the RDR nominal
and low-P options.}
\vspace{-2mm}
\begin{tabular}{l|c|c|c|c|c|c|c|c|c} \hline
& \multicolumn{2}{|c}{Magnets} & \multicolumn{2}{|c}{Diagnostic} &
\multicolumn{5}{|c}{Collimators} \\
& \multicolumn{2}{|c}{} & \multicolumn{2}{|c}{Detectors} &
\multicolumn{5}{|c}{} \\
\cline{2-10}
        & SC &   Warm  & Synch- & Chere- & Energy  & Chere- & Dump    & Dump   &  Dump   \\
        &    &   (max) & rotron & nkov   & chicane & nkov   &  1      &  2     &   3     \\
\hline
TDR     & 0  &  12 W/m & 30 W   & 130 W  &  22 W  & 0.3 kW & 2.8 kW  & 1.9 kW & 3.0 kW  \\
Nominal & 0  & 0.6 W/m & 4 W    &  26 W  &   2 W  & 37 W   & 0.3 kW  & 0.2 kW & 0.2 kW  \\
Low-P   & 0  & 130 W/m & 0.5 kW & 0.6 kW & 0.4 kW & 2.2 kW & 14.8 kW & 9.0 kW & 10.3 kW \\
\hline
\end{tabular}
\normalsize
\label{tab:summary}
\vspace{-2mm}
\end{center}
\end{table}

However, the presented calculations were done assuming ideal collision
conditions. Non-ideal conditions, such as large vertical beam-to-beam
separation at IP, will increase the beam disruption and, consequently,
the extraction beam losses \cite{offset}. Evaluation of these effects
requires a separate study.

\end{document}